\newcommand{\CC}{\Lambda}

\newcommand{\Omo}{\Omega_{m}}

\newcommand{\OLo}{\Omega_{\Lambda}}

\newcommand{\rDE}{\rho_{\rm DE}}

\newcommand{\nueff}{\nu_{\rm eff}}

\newcommand{\oD}{\omega_{\rm BD}}

\newcommand{\dpsi}{\dot{\psi}}
\newcommand{\ddpsi}{\ddot{\psi}}
\newcommand{\wBD}{\omega_{\rm BD}}

\newcommand{\fracdpsipsi}{\frac{\dpsi}{\psi}}

\newcommand{\eBD}{\epsilon_{\rm BD}}
\newcommand{\mPl}{m_{\rm Pl}}

\newcommand{\Geff}{G_{\rm eff}}

\newcommand{\weff}{w_{\rm eff}}

\newcommand{\rv}{\rho_{\rm vac}}
\newcommand{\rvo}{\rho^0_{\rm vac}}

\newcommand{\astar}{a_{*}}
\newcommand{\zstar}{z_{*}}

\documentclass{ws-procs961x669}            
\begin{document}
\title{BD-$\Lambda$CDM and Running Vacuum Models: Theoretical background and current observational status}

\author{Javier de Cruz P\'erez$^1$\footnote{Speaker}, Joan Sol\`a Peracaula$^2$, Adri\`a G\'omez-Valent$^3$  and Cristian Moreno-Pulido$^2$}

\address{$1$ Department of Physics, Kansas State University, 116 Cardwell Hall, Manhattan, KS 66506, USA\\ $2$ Departament de F\'isica Qu\`antica i Astrof\'isica, and Institute of Cosmos Sciences,\\ Universitat de Barcelona,
Av. Diagonal 647, E-08028 Barcelona, Catalonia, Spain\\
$3$ Dipartamento di Fisica, and INFN Sezione di Roma Tor Vergata, via della Ricerca Scientifica 1, 00133 Roma Italy\\
$^*$E-mail: decruz@fqa.ub.edu, sola@fqa.ub.edu, cristian.moreno@fqa.ub.edu, agvalent@roma2.infn.it}

\begin{abstract}
We present an analysis of the Brans-Dicke cosmological model with a cosmological constant and cold dark matter (BD-$\Lambda${CDM}). We find that the BD-$\Lambda${CDM} is favored by the overall cosmological data (SNIa+BAO+$H(z)$+LSS+CMB) when it is compared with the standard model of cosmology. The BD-$\Lambda$CDM model can be viewed from the GR perspective as a Running Vacuum Model (RVM) with a time evolving vacuum energy density. Due to this fact and also to its time evolving effective gravitational coupling, the model can alleviate the $\sigma_8$ and the $H_0$ tensions at a time.  We also present the results for different types of RVM's when they are tested in the light of the cosmological data and we show that a mild dynamics for the vacuum energy density can help to smooth out the aforementioned tensions, thus improving the performance of $\Lambda$CDM model. 
\end{abstract}


\bodymatter

\section{Introduction}\label{Sec:Introduction}
The high quality observations performed during the last two decades, have allowed to demonstrate, with a high confidence range, that the Universe is in expansion and to be more precise in accelerated expansion \cite{Riess:1998cb, Perlmutter:1998np}. The standard model of cosmology, the so-called $\Lambda$CDM, is based on the assumption that the observed accelerated expansion is due to the existence of a repulsive force, exerted by the $\Lambda$ term, which counteracts the attractive gravitational force and pushes the clusters of galaxies at a speed that increases with the cosmic expansion. This term, which is called the cosmological constant (CC), is associated with a mysterious form of energy, usually called dark energy (DE) presumably permeating all corners of the Universe as a whole. The cosmological constant is not the only element beyond the conventional matter, namely: baryons, photons and neutrinos, demanded by the observations since it is also essential the presence of large amounts of what is commonly call, cold dark matter (CDM). It is important to remark that the concordance model is defined in the framework of General Relativity (GR), whose field equations were found by Einstein in 1915 \cite{Einstein:1915ca}. The fact that the $\Lambda$CDM model, with so few ingredients (encoded in 6 degrees of freedom), has remained robust and unbeaten for a long time turns out to be pretty impressive. It is consistent with a large body of very accurate cosmological data, like the cosmic microwave background (CMB), the baryonic acoustic oscillations (BAO) or the large scale structure (LSS) data.  
\newline
However, this does not mean at all that we really understand the primary (dynamical) cause for such an acceleration or that we are in position to propose a cosmological model explaining  the speeding up of the cosmos at the level of fundamental physics, say quantum field theory (QFT) in curved spacetime, quantum gravity or string theory. For instance, if we try to explain the origin of the cosmological constant in the context of QFT we will immediately realize that we have to face the well-known old cosmological constant problem. This problem was formulated by Zel'dovich \cite{Zeldovich:1967,Zeldovich:1968} (see also \cite{Weinberg:1988cp,Peebles:2002gy,Padmanabhan:2002ji,Copeland:2006wr}) and basically consists in the tremendous mismatch between the computed value of the vacuum energy density, considering different contributions from QFT, and the experimental measure, being the discrepancy between both values in natural units of at least 55 orders of magnitude. 
\newline
Thus, in spite of the many virtues of the $\Lambda$CDM model, the lack of an explanation for the presence of the CC, leads inevitably to an unsatisfactory theoretical picture which has motivated the search of a wide range of alternatives beyond the standard model. What is more, aside from the theoretical problems some persisting tensions with the cosmological data (being the most important ones, the tension on the local value of the $H_0$ parameter \cite{Planck:2018vyg,Reid:2019tiq} and the one affecting the LSS data \cite{Macaulay:2013swa}) point out that the $\Lambda$CDM model might be performing insufficiently at the observational level. 
\newline
Here we study two alternatives where the aforementioned tension could be alleviated. It has been suggested that it would help to alleviate the tensions if DE would be a dynamical quantity (see for instance the possibility of an early dark energy \cite{Gomez-Valent:2021cbe,Benevento:2020fev}), i.e. slowly evolving with the cosmic expansion. This could be achieved, for example, through scalar field models \cite{Peebles:1987ek,Ratra:1987rm,Peccei:1987mm,Wetterich:1987fm,Wetterich:1994bg,Amendola:1999er} (see \cite{Gomez-Valent:2020mqn} for an updated analysis), where the energy density associated to the $\Lambda$ term is replaced by the energy density of the scalar field, which varies throughout the cosmic history. In this paper, instead of considering a scalar field model we focus on another promising possibility, namely that vacuum energy density $\rho_{\rm vac} = \Lambda/8\pi{G_N}$ might be a time evolving quantity whose dynamics is triggered by the quantum effects coming from the matter fields. We call these types of models Running Vacuum Models (RVM's). Phenomenologically, these models, have recently been carefully confronted against the wealth of the cosmological data, obtaining a significant success \cite{Sola:2016jky,SolaPeracaula:2017esw,Sola:2017znb,Gomez-Valent:2018nib,Sola:2018sjf,SolaPeracaula:2021gxi}. 
\newline
On the other hand there exists the possibility to consider a completely different approach, where a theoretical framework, different from GR, is assumed. In this regard, Brans \& Dicke \cite{BransDicke1961} proposed the first historical attempt to extend Einstein's GR by promoting the Newtonian coupling $G_N$ into a variable one in the cosmic time $G(t)$. Gravity is then not only mediated by the metric but also by an scalar field denoted by $\psi$. The theory contains an extra {\it d.o.f.} encoded in $\psi$ and a dimensionless parameter $\omega_{\rm BD}$. In order to recover GR a necessary (though not sufficient) condition is to demand large values of $\omega_{\rm BD}$. We consider the presence of the CC and all the species considered in the standard model, so we call the model studied in the context of Brans-Dicke (BD) gravity, BD-$\Lambda$CDM. 
\newline 
In this work we present a summary of the main results recently obtained by confronting the overall cosmological observations with the BD-$\Lambda$CDM model and related RVM's and we show how they deal with the aforementioned tensions. See the original references \cite{SolaPeracaula:2018wwm,SolaPeracaula:2019zsl,SolaPeracaula:2020vpg,SolaPeracaula:2021gxi} for more details. 
\section{Background equations for the BD-$\Lambda$CDM model}
We will consider the original BD-action extended with a cosmological constant density term as it is essential to mimic the conventional $\CC$CDM model based on GR and reproduce its main successes. We call this the `BD-$\CC$CDM model', i.e. the version of the  $\CC$CDM within the BD paradigm.
The BD action reads,  written in the Jordan frame, as follows\,\footnote{We use natural units, $\hbar=c=1$ and $G_N=1/\mPl^2$, where $\mPl\simeq 1.22\times 10^{19}$ GeV is the Planck mass. As for the geometrical quantities, we adopt the $(+, +, +)$ convention of the popular classification by Misner, Thorn and Wheeler\,\cite{MTW:1974}.}:
\begin{eqnarray}
S_{\rm BD}=\int d^{4}x\sqrt{-g}\left[\frac{1}{16\pi}\left(R\psi-\frac{\oD}{\psi}g^{\mu\nu}\partial_{\nu}\psi\partial_{\mu}\psi\right)-\rho^{0}_{\rm vac}\right]+S_m\,. \label{eq:BDaction}
\end{eqnarray}
For the sake of convenience we denote the constant vacuum energy density as $\rho^0_{\rm vac}$. The (dimensionless) factor  in front of the kinetic term of $\psi$, i.e. $\oD$, will be referred to as the BD-parameter and we consider the canonical option, $\oD=$const.
The last term of (\ref{eq:BDaction}) stands for the matter action $S_{m}$, which is constructed from the Lagrangian density of the matter fields. There is no potential for the BD-field $\psi$ in the original BD-theory, but we admit the presence of a CC term associated to $\rho^{0}_{\rm vac}$. \newline
\newline
The modified field equations (with respect to GR) can be obtained after performing variation of the action \eqref{eq:BDaction}  with respect to both the metric and the scalar field $\psi$.  While the  first variation yields
\begin{equation}\label{eq:BDFieldEquation1}
\psi\,G_{\mu\nu}+\left(\Box\psi +\frac{\oD}{2\psi}\left(\nabla\psi\right)^2\right)\,g_{\mu\nu}-\nabla_{\mu}\nabla_{\nu}\psi-\frac{\oD}{\psi}\nabla_{\mu}\psi\nabla_{\nu}\psi=8\pi\left(\,T_{\mu\nu}-g_{\mu\nu}\rho^{0}_{\rm vac}\right)\,,
\end{equation}
the second variation gives the wave equation for $\psi$
\begin{equation}\label{eq:BDFieldEquation2}
\Box\psi=\frac{8\pi}{2\oD+3}\,\left(T-4\rho^{0}_{\rm vac}\right)\,.
\end{equation}
We have used the definition $(\nabla\psi)^2\equiv g^{\mu\nu}\nabla_{\mu}\psi\nabla_{\nu}\psi$.
In the first equation, $G_{\mu\nu}=R_{\mu\nu}-(1/2)Rg_{\mu\nu}$ is the Einstein tensor, and  on its {\it r.h.s.} $T_{\mu \nu}=-(2/\sqrt{-g})\delta S_{m}/\delta g^{\mu\nu}$ is the energy-momentum tensor of matter, being 
 $T \equiv T^\mu_\mu$ its trace. The total energy-momentum tensor is the sum of the matter and vacuum contributions and it takes the perfect fluid form:
\begin{equation} \label{eq:EMT}
\tilde{T}_{\mu\nu} =T_{\mu\nu}  -\rho^{0}_{\rm vac} g_{\mu\nu}=p\,g_{\mu\nu}+(\rho + p)u_\mu{u_\nu}\,,
\end{equation}
{with $\rho \equiv \rho_m + \rho_\gamma+\rho_{\rm ncdm} + \rho^{0}_{\rm vac}$ and $p \equiv p_m + p_\gamma + p_{\rm ncdm} + p^{0}_{\rm vac}$. The matter part $\rho_m \equiv \rho_b + \rho_{\rm cdm}$, contains the pressureless contribution from baryons and cold dark matter and photons are of course relativistic, so $p_\gamma=\rho_\gamma/3$. The functions $\rho_{\rm ncdm}$ and $p_{\rm ncdm}$ (`${\rm ncdm}$' means non-CDM) include the effect of massive and massless neutrinos, and therefore must be computed numerically.} 
As in GR, we have considered a constant vacuum energy density, $\rho^{0}_{\rm vac}$, in the BD-action (\ref{eq:BDaction}), being its equation of state $p^{0}_{\rm vac}=-\rho^{0}_{\rm vac}$ the usual one. 
\newline
\newline
Let us write down the field equations in the flat (to know more about the nonflat models see the references \cite{Ooba:2017ukj,Ooba:2017lng,Park:2018bwy,Park:2018fxx,Park:2018tgj,Park:2019emi,Ryan:2019uor,Khadka:2020vlh,Khadka:2020whe,Cao:2020evz,Cao:2021ldv,Khadka:2021xcc,Cao:2021cix}) FLRW metric,  $ds^2=-dt^2 + a^2\delta_{ij}dx^idx^j$. Using the previous definitions for the total energy density $\rho$ and pressure $p$ we get the two independent equations:
\begin{equation}
3H^2 + 3H\fracdpsipsi -\frac{\wBD}{2}\left(\fracdpsipsi\right)^2 = \frac{8\pi}{\psi}\rho\label{eq:Friedmannequation}
\end{equation}
and
\begin{equation}
2\dot{H} + 3H^2 + \frac{\ddpsi}{\psi} + 2H\frac{\dpsi}{\psi} + \frac{\wBD}{2}\left(\frac{\dpsi}{\psi}\right)^2 = -\frac{8\pi}{\psi}p\,,\label{eq:pressureequation}
\end{equation}
whereas from (\ref{eq:BDFieldEquation2}) we obtain
\begin{equation}\label{eq:FieldeqPsi}
\ddpsi +3H\dpsi = \frac{8\pi}{2\wBD +3}(\rho - 3p)\,.
\end{equation}
Here dots stand for derivatives with respect to the cosmic time and $H=\dot{a}/a$ is the Hubble rate. In the limit $\psi \rightarrow 1/G_N$ and $\oD\to\infty$ we recover the standard equations. The connection between GR and  the $\oD\to\infty$ limit is sometimes not as straightforward as one might  naively think\,\cite{Faraoni:1998yq,Faraoni:1999yp}. 
\newline
\newline
Matter and the BD-field are not in interaction, as a result, the local conservation law adopts the usual form:
\begin{equation}\label{eq:FullConservationLaw}
\dot{\rho} + 3H(\rho + p)=\sum_{N} \left[\dot{\rho}_N + 3H(\rho_N + p_N)\right] = 0\,,
\end{equation}
where the sum is over all components, i.e. baryons, dark matter, neutrinos, photons and vacuum. We assume that all of the components are separately conserved in the main periods of the cosmic evolution.
For convenience, we will use a dimensionless BD-field, $\varphi$, and the inverse of the BD-parameter, according to :
\begin{equation}\label{eq:definitions}
\varphi(t) \equiv G_N\psi(t)\,,\qquad  \qquad\epsilon_{\rm BD} \equiv \frac{1}{\omega_{\rm BD}}\,.
\end{equation}
As stated before $G_N$ gives the local value of the gravitational coupling. Note that a nonvanishing value of $\eBD$ entails a deviation from GR. Being $\varphi(t)$  a dimensionless quantity,  we can recover GR by enforcing the simultaneous limits   $\epsilon_{\rm BD} \rightarrow 0$ \textit{and} $\varphi\to 1 $.   

\renewcommand{\arraystretch}{0.7}
\begin{table*}[t!]
\begin{center}
\resizebox{1\textwidth}{!}{
\begin{tabular}{|c  |c | c |  c | c | c  |c  |}
 \multicolumn{1}{c}{} & \multicolumn{4}{c}{Baseline}
\\\hline
{\scriptsize Parameter} & {\scriptsize $\Lambda$CDM}  & {\scriptsize type I RRVM} & {\scriptsize type I RRVM$_{\rm thr.}$}  &  {\scriptsize type II RRVM} &  {\scriptsize BD-$\Lambda$CDM}
\\\hline
{\scriptsize $H_0$(km/s/Mpc)}  & {\scriptsize $68.37^{+0.38}_{-0.41}$} & {\scriptsize $68.17^{+0.50}_{-0.48}$} & {\scriptsize $67.63^{+0.42}_{-0.43}$}  & {\scriptsize $69.02^{+1.16}_{-1.21}$} & {\scriptsize $69.30^{+1.38}_{-1.33}$}
\\\hline
{\scriptsize$\omega_b$} & {\scriptsize $0.02230^{+0.00019}_{-0.00018}$}  & {\scriptsize $0.02239^{+0.00023}_{-0.00024}$} & {\scriptsize $0.02231^{+0.00020}_{-0.00019}$}  &  {\scriptsize $0.02245^{+0.00025}_{-0.00027}$} & {\scriptsize $0.02248\pm 0.00025$}
\\\hline
{\scriptsize$\omega_{\rm cdm}$} & {\scriptsize $0.11725^{+0.00094}_{-0.00084}$}  & {\scriptsize $0.11731^{+0.00092}_{-0.00087}$} & {\scriptsize $0.12461^{+0.00201}_{-0.00210}$}  &  {\scriptsize $0.11653^{+0.00158}_{-0.00160}$} &   {\scriptsize $0.11629^{+0.00148}_{-0.00151}$}
\\\hline
{\scriptsize$\nu_{\rm eff}$} & {-}  & {\scriptsize $0.00024^{+0.00039}_{-0.00040}$} & {\scriptsize $0.02369^{+0.00625}_{-0.00563}$}  &  {\scriptsize $0.00029\pm 0.00047$} & {-}
\\\hline
{\scriptsize$\epsilon_{\rm BD}$} & {-}  & {-} & {-}  &  {-} & {\scriptsize $-0.00109\pm ^{+0.00135}_{-0.00141}$}
\\\hline
{\scriptsize$\varphi_{\rm ini}$} & {-}  & {-} & {-}  &  {\scriptsize $0.980^{+0.031}_{-0.027}$} & {\scriptsize $0.972^{+0.030}_{-0.037}$}
\\\hline
{\scriptsize$\varphi_0$} & {-}  & {-} & {-}  &  {\scriptsize $0.973^{+0.036}_{-0.033}$} & {\scriptsize $0.963^{+0.036}_{-0.041}$}
\\\hline
{\scriptsize$\tau_{\rm reio}$} & {{\scriptsize$0.049^{+0.008}_{-0.007}$}} & {{\scriptsize$0.051^{+0.008}_{-0.009}$}} & {{\scriptsize$0.058^{+0.007}_{-0.009}$}}  &   {{\scriptsize$0.051\pm 0.008$}} & {{\scriptsize$0.051\pm 0.008$}}
\\\hline
{\scriptsize$n_s$} & {{\scriptsize$0.9698^{+0.0039}_{-0.0036}$}}  & {{\scriptsize$0.9716^{+0.0044}_{-0.0047}$}} & {{\scriptsize$0.9703\pm 0.038$}} &   {{\scriptsize$0.9762^{+0.0081}_{-0.0091}$}} &
\\\hline
{\scriptsize$\sigma_8$}  & {{\scriptsize$0.796\pm 0.007$}}  & {{\scriptsize$0.789^{+0.013}_{-0.014}$}} & {{\scriptsize$0.768^{+0.010}_{-0.009}$}}  &   {{\scriptsize$0.791^{+0.013}_{-0.012}$}} & {{\scriptsize$0.790^{+0.013}_{-0.012}$}}
\\\hline
{\scriptsize$S_8$}  & {{\scriptsize$0.796\pm 0.011$}}  & {{\scriptsize$0.791^{+0.014}_{-0.013}$}} & {{\scriptsize$0.797^{+0.012}_{-0.011}$}}  &   {{\scriptsize$0.781^{+0.021}_{-0.020}$}} & {{\scriptsize$0.777^{+0.021}_{-0.022}$}}
\\\hline
{\scriptsize$r_s$ (Mpc)}  & {{\scriptsize$147.90^{+0.30}_{-0.31}$}}  & {{\scriptsize$147.99^{+0.35}_{-0.36}$}} & {{\scriptsize$147.81\pm 0.30$}}  &   {{\scriptsize$146.30^{+2.39}_{-2.30}$}} & {{\scriptsize$145.72^{+2.44}_{-2.90}$}}
\\\hline
{\scriptsize$\chi^2_{\rm min}$}  & {{\scriptsize 2290.20}}  & {{\scriptsize 2289.72}} & {{\scriptsize 2272.44}}  &   {{\scriptsize 2288.74}} & {{\scriptsize 2289.40}}
\\\hline
{\scriptsize$\Delta{\rm DIC}$}  & {-}  & {{\scriptsize -2.70}} & {{\scriptsize +13.82}}  &   {{\scriptsize -4.59}} & {{\scriptsize -3.53}}
\\\hline
\end{tabular}}
\end{center} 
\begin{tabnote}
$^{}$ {\scriptsize The mean values and 68.3\% confidence limits for the models under study using our Baseline dataset, which is comporsed by: the full Pantheon likelihood \cite{Scolnic:2017caz}, 13 BAO data points \cite{Carter:2018vce,Kazin:2014qga,Gil-Marin:2016wya,Abbott:2017wcz,Neveux:2020voa,duMasdesBourboux:2020pck}, 31 data points on $H(z_i)$, at different redshifts \cite{Jimenez:2003iv,Simon:2004tf,Stern:2009ep,Moresco:2012jh,Zhang:2012mp,Moresco:2015cya,Moresco:2016mzx,Ratsimbazafy:2017vga}, 14 points on the observable $f(z_i)\sigma_8(z_i)$ \cite{Said:2020epb,Simpson:2015yfa,Blake:2013nif,Gil-Marin:2016wya,Mohammad:2018mdy,Okumura:2015lvp,Neveux:2020voa} and finally the full Planck likelihood for the CMB data \cite{Planck:2018vyg}. We display the fitting values for the usual parameters, to wit: $H_0$, the reduced density parameter for baryons ($w_b = \Omega^0_b{h^2}$) and also for CDM ($w_{\rm cdm} = \Omega^0_{\rm cdm}{h^2}$), being $\Omega_i^0=8\pi G_N\rho^0_i/3H_0^2$ and $h$ the reduced Hubble constant, the reionization optical depth $\tau_{\rm reio}$, the spectral index $n_s$ and the current matter density rms fluctuations within spheres of radius $8h^{-1}$~Mpc, i.e.  $\sigma_8$. We include also a couple of derived parameters, namely: the sound horizon at the baryon drag epoch $r_s$ and $S_8\equiv \sigma_8\sqrt{\Omega^0_m/0.3}$. For the RRVM's we provide the value of $\nu_{\rm eff}$, and for the type II and BD-$\Lambda$CDM \cite{SolaPeracaula:2021gxi} we also report the initial and current values of $\varphi$, $\varphi_{\rm ini}$ and $\varphi_0$, respectively. {The parameter $\epsilon_{\rm BD}\equiv1/\omega_{\rm BD}$ (inverse of the Brans-Dicke parameter\,\cite{BransDicke1961}) controls the dynamics of the scalar field\cite{SolaPeracaula:2019zsl,SolaPeracaula:2020vpg}.
We finally provide the corresponding values of $\chi^2_{\rm min}$ and $\Delta$DIC.}}\\
\end{tabnote}\label{aba:tbl1}
\label{tableFit1}
\end{table*}
%

\section{Connection of the BD-$\Lambda$CDM model with the Running Vacuum Model}
So far, no analytical solutions to the system \eqref{eq:Friedmannequation}-\eqref{eq:FieldeqPsi} have been found, for this reason our actual analysis proceeds numerically. However, it is possible to search for approximate solutions valid in the different epochs of cosmic history, which can provide a qualitative understanding of the numerical results obtained. Actually, a first attempt in this direction trying to show that BD-$\CC$CDM  can mimic the Running Vacuum Model (RVM) was done in \cite{Peracaula:2018dkg,Perez:2018qgw}, we refer the reader to these references for details. See also \cite{Banerjee:2000mj,Banerjee:2000gt}. We are interested in looking for  solutions in the Matter Dominated Epoch (MDE) in the form of a power-law ansatz in which the BD-field $\varphi$ evolves very slowly:
\begin{equation}
\varphi(a) = \varphi_0\,a^{-\epsilon}\  \qquad (|\epsilon|\ll1)\,.\label{powerlaw}
\end{equation}
The $\epsilon$ parameter must be a very small parameter in absolute value since $G(a)\equiv G(\varphi(a))$ cannot depart too much from $G_N$. For $\epsilon>0$, the effective coupling increases with the expansion and hence is asymptotically free since  $G(a)$ is smaller in the past, which is the epoch when the
Hubble rate (with natural dimension of energy) is bigger. For  $\epsilon<0$, is the other way around and $G(a)$ decreases with the expansion.
\newline
\newline
Plugging the power-law ansatz into the cosmological equations we end up with the following pair of Friedmann-like equations to ${\cal O}(\epsilon)$\cite{Peracaula:2018dkg,Perez:2018qgw}:
\begin{equation}\label{eq:effective Friedmann}
   H^2=\frac{8\pi G}{3}\left(\rho^0_{m} a^{-3+\epsilon}+\rDE(H)\right)
\end{equation}
and for the acceleration equation
\begin{equation}\label{eq:currentacceleration}
\frac{\ddot{a}}{a}=-\frac{4\pi G}{3}\,\left(\rho_m^0 a^{-3+\epsilon}+\rDE(H)+3p^{0}_{\rm vac}\right)\,,
\end{equation}
with $G=G_N/\varphi_0$. The first of the above equations can be understood as an effective Friedmann's equation with time-evolving cosmological term, in which the DE appears as if it were a dynamical quantity:
\begin{equation}\label{eq:rLeff}
  \rDE(H)=\rho^{0}_{\rm vac}+\frac{3\,\bar{\nu}_{\rm eff}}{8\pi G} H^2\,.
\end{equation}
Here
\begin{equation}
\ \bar{\nu}_{\rm eff}\equiv\epsilon\left(1+\frac16\,\oD\epsilon\right) \label{nueff}
\end{equation}
is the coefficient controlling the dynamical character of the effective dark energy \eqref{eq:rLeff}. The structure of this dynamical dark energy  (DDE) is reminiscent of the Running Vacuum Model (RVM), see \cite{Sola:2013gha,Sola:2015rra,Gomez-Valent:2017tkh} and references therein. 
Notice from \eqref{eq:effective Friedmann} that,  to ${\cal O}(\epsilon)$:
\begin{equation}\label{eq:SumRule}
\Omo +\OLo =1-\bar{\nu}_{\rm eff}\,,
\end{equation}
so, due to the presence of $\bar{\nu}_{\rm eff}$, we find a slight deviation from the usual sum rule of GR. 
Only in the case $\epsilon=0$, we have $\bar{\nu}_{\rm eff}=0$ and then  we recover the usual cosmic sum rule. 
As stated, the parameter $\bar{\nu}_{\rm eff}$ becomes associated to the dynamics of the dark energy. Interestingly, the above expression \eqref{eq:rLeff} adopts the form of the RVM, see  \cite{Sola:2013gha,Sola:2015rra,Gomez-Valent:2017tkh} and references therein. We shall see more about the most important features of the RVM in the next section. All in all, it turns out that the BD-RVM (we may call it in this way for convenience) can cure the $H_0$-tension due to the evolution of the effective gravitational $\Geff$ and, additionally, by mimicking the RVM it can also alleviate the well-known $\sigma_8$-tension thanks to the role played by $\bar{\nu}_{\rm eff}$. 
Going a little further in this effective picture from \eqref{eq:currentacceleration} we can define the EoS parameter for the effective DDE: 
\begin{equation}\label{eq:EffEoS}
\weff(z)=\frac{p^{0}_{\rm vac}}{\rDE(H)}\simeq -1+\frac{3\bar{\nu}_{\rm eff}}{8\pi G \rho^{0}_{\rm vac}}\,H^2(z)=-1+\frac{\bar{\nu}_{\rm eff}}{\Omega_\CC}\,\frac{H^2(z)}{H_0^2}\,,
\end{equation}
where use has been made of (\ref{eq:rLeff}). As it is clear from the above equation the BD-RVM, unlike the original RVM, does not describe a dark energy of pure vacuum form but a DE whose EoS departs, in a mild way, from the pure vacuum. Actually, for $\epsilon>0\ (\epsilon<0) $  we have $\bar{\nu}_{\rm eff}>0\  (\bar{\nu}_{\rm eff}<0)$ and the effective DDE behaves quintessence (phantom)-like. 

\renewcommand{\arraystretch}{0.65}
\begin{table*}[t!]
\begin{center}
\resizebox{1\textwidth}{!}{
\begin{tabular}{|c  |c | c |  c | c | c  |c  |}
 \multicolumn{1}{c}{} & \multicolumn{4}{c}{Baseline + $H_0$}
\\\hline
{\scriptsize Parameter} & {\scriptsize $\Lambda$CDM}  & {\scriptsize type I RRVM} & {\scriptsize type I RRVM$_{\rm thr.}$}  &  {\scriptsize type II RRVM} &  {\scriptsize BD-$\Lambda$CDM}
\\\hline
{\scriptsize $H_0$ (km/s/Mpc)}  & {\scriptsize $68.75^{+0.41}_{-0.36}$} & {\scriptsize $68.77^{+0.49}_{-0.48}$} & {\scriptsize $68.14^{+0.43}_{-0.41}$}  & {\scriptsize $70.93^{+0.93}_{-0.87}$}  & {\scriptsize $71.23^{+1.01}_{-1.02}$}
\\\hline
{\scriptsize$\omega_b$} & {\scriptsize $0.02240^{+0.00019}_{-0.00021}$}  & {\scriptsize $0.02238^{+0.00021}_{-0.00023}$} & {\scriptsize $0.02243^{+0.00019}_{-0.00018}$}  &  {\scriptsize $0.02269^{+0.00025}_{-0.00024} $}  & {\scriptsize $0.02267^{+0.00026}_{-0.00023} $}
\\\hline
{\scriptsize$\omega_{\rm cdm}$} & {\scriptsize $0.11658^{+0.00080}_{-0.00083}$}  & {\scriptsize $0.11661^{+0.00084}_{-0.00085}$} & {\scriptsize $0.12299^{+0.00197}_{-0.00203}$}  &  {\scriptsize $0.11602^{+0.00162}_{-0.00163}$}  & {\scriptsize $0.11601^{+0.00161}_{-0.00157}$}
\\\hline
{\scriptsize$\nu_{\rm eff}$} & {-}  & {\scriptsize $-0.00005^{+0.00040}_{-0.00038}$} & {\scriptsize $0.02089^{+0.00553}_{-0.00593}$}  &  {\scriptsize $0.00038^{+0.00041}_{-0.00044}$}  & {-}
\\\hline 
{\scriptsize$\epsilon_{\rm BD}$} & {-}  & {-} & {-}  &  {-} & {\scriptsize $-0.00130\pm ^{+0.00136}_{-0.00140}$}
\\\hline
{\scriptsize$\varphi_{\rm ini}$} & {-}  & {-} & {-}  &  {\scriptsize $0.938^{+0.018}_{-0.024}$}  & {\scriptsize $0.928^{+0.024}_{-0.026}$}
\\\hline
{\scriptsize$\varphi_0$} & {-}  & {-} & {-}  &  {\scriptsize $0.930^{+0.022}_{-0.029}$}  & {\scriptsize $0.919^{+0.028}_{-0.033}$}
\\\hline
{\scriptsize$\tau_{\rm reio}$} & {{\scriptsize$0.050^{+0.008}_{-0.007}$}} & {{\scriptsize$0.049^{+0.009}_{-0.008}$}} & {{\scriptsize$0.058^{+0.008}_{-0.009}$}}  &   {{\scriptsize$0.052\pm 0.008$}}  & {{\scriptsize$0.052\pm 0.008$}} 
\\\hline
{\scriptsize$n_s$} & {{\scriptsize$0.9718^{+0.0035}_{-0.0038}$}}  & {{\scriptsize$0.9714\pm 0.0046$}} & {{\scriptsize$0.9723^{+0.0040}_{-0.0039}$}} &   {{\scriptsize$0.9868^{+0.0072}_{-0.0074}$}}  & {{\scriptsize$0.9859^{+0.0073}_{-0.0072}$}}
\\\hline
{\scriptsize$\sigma_8$}  & {{\scriptsize$0.794\pm 0.007$}}  & {{\scriptsize$0.795\pm 0.013$}} & {{\scriptsize$0.770\pm 0.010$}}  &   {{\scriptsize$0.794^{+0.013}_{-0.012}$}}  & {{\scriptsize$0.792^{+0.013}_{-0.012}$}} 
\\\hline
{\scriptsize$S_8$}  & {{\scriptsize$0.788^{+0.010}_{-0.011}$}}  & {{\scriptsize$0.789\pm 0.013$}} & {{\scriptsize$0.789\pm 0.011$}}  &   {{\scriptsize$0.761^{+0.018}_{-0.017}$}}  & {{\scriptsize$0.758^{+0.019}_{-0.018}$}}
\\\hline
{\scriptsize$r_s$ (Mpc)}  & {{\scriptsize$147.97^{+0.29}_{-0.31}$}}  & {{\scriptsize$147.94^{+0.35}_{-0.36}$}} & {{\scriptsize$147.88^{+0.33}_{-0.29}$}}  &   {{\scriptsize$143.00^{+1.54}_{-1.96}$}}  & {{\scriptsize$142.24^{+1.99}_{-2.12}$}}
\\\hline
{\scriptsize$\chi^2_{\rm min}$}  & {{\scriptsize 2302.14}}  & {{\scriptsize 2301.90}} & {{\scriptsize 2288.82}}  &   {{\scriptsize 2296.38}}  & {{\scriptsize 2295.36}}
\\\hline
{\scriptsize$\Delta{\rm DIC}$}  & {-}  & {{\scriptsize -2.36}} & {{\scriptsize +10.88}}  &   {{\scriptsize +5.52}}  & {{\scriptsize +6.25}}
\\\hline
\end{tabular}}
\end{center}
\begin{tabnote}
$^{}$ {\scriptsize Same as in Table 1, but also considering the prior on $H_0=(73.5\pm 1.4)$ km/s/Mpc from SH0ES \cite{Reid:2019tiq}.}\\
\end{tabnote}\label{aba:tbl1}
\label{tableFit2}
\end{table*}
%

\section{Background equations for the RVM's }
In the following we are going to present the theoretical framework where the previously mentioned RVM's are placed \cite{Sola:2013gha,Sola:2015rra,Gomez-Valent:2017tkh} and references therein. We consider two types of dynamical vacuum energy (DVE) scenarios.  In type I scenario the vacuum is in interaction with matter, in contrast, in type II  matter is conserved at the expense of an exchange between the vacuum and a slowly evolving gravitational coupling $G (H)$. In both cases, the combined cosmological `running' of these quantities  insures the accomplishment of the Bianchi identity (and the  local conservation law).
Let us therefore consider a generic cosmological framework described by the spatially flat Friedmann-Lema\^\i tre-Robertson-Walker (FLRW) metric. The vacuum energy density in the RVM can be expressed as \,\cite{Sola:2013gha, Sola:2015rra}:
\begin{equation}\label{eq:RVMvacuumdadensity}
\rv(H) = \frac{3}{8\pi{G}_N}\left(c_{0} + \nu{H^2+\tilde{\nu}\dot{H}}\right)+{\cal O}(H^4)\,.
\end{equation}
The ${\cal O}(H^4)$ terms play no role in the post-inflationary epoch so they can be neglected. 
The expression \eqref{eq:RVMvacuumdadensity} can be motivated from the explicit QFT calculations on a FLRW background\,\cite{Moreno-Pulido:2020anb}.
The value of the additive constant $c_0$ is fixed by the boundary condition $\rho_{\rm vac}(H_0)=\rvo$. The two dynamical components $H^2$ and $\dot{H}$ are dimensionally homogeneous and in principle independent.
The dimensionless coefficients $\nu$ and $\tilde{\nu}$ encode the dynamics of the vacuum at low energy and we naturally expect for both of them  $|\nu,\tilde{\nu}|\ll1$. An estimate of $\nu$ in QFT indicates that it is of order $10^{-3}$ at most \cite{Sola:2007sv}. In the calculations performed in \cite{Moreno-Pulido:2020anb} these coefficients are expected to be of order $\sim M_X^2/\mPl^2\ll 1$, being $M_X$ of order of a typical Grand Unified Theory (GUT) scale. Taking into account the multiplicity of particles in a GUT, the theoretical estimate on $\nu$ could be much larger but still subject to a value below 1. 
\newline
\newline
We are interested in a particular form for the vacuum energy density obtained by imposing $\tilde{\nu}=\nu/2$. Consequently
\begin{equation}
\rv(H) ={3}/(8\pi G_N)\left[c_0 + \nu\left({H^2+\frac12\dot{H}}\right)\right].    
\end{equation}
We will call this form of the vacuum energy density the `RRVM' since it realizes the generic RVM density \eqref{eq:RVMvacuumdadensity} through the Ricci scalar $\mathcal{R} = 12H^2 + 6\dot{H}$, namely
\begin{equation}\label{RRVM}
\rv(H) =\frac{3}{8\pi{G_N}}\left(c_0 + \frac{\nu}{12}\mathcal{R}\right)\equiv \rv(\mathcal{R})\,.
\end{equation}
Due to the fact that the condition $\mathcal{R}/H^2\ll 1$ is fulfilled in the early epochs of the cosmological evolution we do not generate any conflict with the BBN nor with any other feature of the modern universe. Of course, early on the RVM has its own mechanism for inflation see \cite{Sola:2013gha,Sola:2015rra,Lima:2013dmf, Sola:2015csa} for more details. %
\subsection{\rm Type I RRVM}
Taking into account, the dynamical character presented for the vacuum energy density, the Friedmann and the acceleration equations for the different species involved, read
\begin{align}
3H^2 &= 8\pi{G_N}\left(\rho_m + \rho_{\rm{ncdm}} + \rho_\gamma + \rv(H)\right),\label{FriedmannEquation} \\
3H^2 + 2\dot{H} &= -8\pi{G_N}\left(p_{\rm{ncdm}} + p_\gamma + p_{\rm vac}(H)\right)\label{AccelerationEquation}\,.
\end{align}
As for the BD-model, we define the total nonrelativistic matter density as the sum of the CDM component and the baryonic one: $\rho_m = \rho_{\rm cdm} + \rho_b$. Therefore the total (relativistic and nonrelativistic)  matter density is $\rho_t=\rho_m+ \rho_\gamma+\rho_{\rm ncdm}$. In a similar way, the total matter pressure reads  $p_t=p_{\rm{ncdm}} + p_\gamma$  (with  $p_\gamma=(1/3)\rho_\gamma$).
Making use of the functions employed by the system solver \texttt{CLASS}\,\cite{Blas:2011rf} we distinguish between the contributions of the nonrelativistic neutrinos
$\rho_h = \rho_{\rm ncdm} - 3p_{\rm ncdm}$ and the one from the relativistic neutrinos $\rho_\nu = 3p_{\rm ncdm}$. 
This separation allows to compute  $\mathcal{R}/12=H^2 + (1/2)\dot{H}$ appearing in \eqref{RRVM} in terms of the energy densities and pressures using \eqref{FriedmannEquation} and \eqref{AccelerationEquation}:
\begin{equation}\label{combination}
\mathcal{R} = 8\pi{G_N}\left(\rho_m + 4\rv + \rho_h\right)\,.
\end{equation}
We can safely neglect the contribution of neutrinos from \eqref{combination} since it remains well below the contribution of CDM and baryons throughout the whole cosmic history.  This fact allows us to solve for the vacuum density as a function of the scale factor $a$ as follows:
\begin{equation}\label{eq:VacuumDens}
\rv(a) = \rvo + \frac{\nu}{4(1-\nu)}(\rho_m(a) - \rho^0_m)\,,
\end{equation}
where ` $0$' (used as subscript or superscript) always refers to current quantities. For $a=1$ we confirm the correct normalization:  $\rv(a=1) = \rvo$. Matter does not follow the standard dilution law, remember that it is in interaction with vacuum, therefore $\rho_m(a)$ is not just $\sim a^{-3}$. The local conservation law for CDM and vacuum can be expressed as follows:
\begin{equation}\label{eq:LocalConsLaw}
\dot{\rho}_{\rm cdm} + 3H\rho_{\rm cdm} = -\dot{\rho}_{\rm vac}\,.
\end{equation}
We assume that baryons are self conserved, which implies $\dot{\rho}_b + 3H\rho_b =0$, and as a consequence the total matter contribution ($\rho_m$)  satisfies the same local conservation law \eqref{eq:LocalConsLaw} as CDM: 
$\dot{\rho}_m + 3H\rho_m = -\dot{\rho}_{\rm vac}$.  Using it together with \eqref{eq:VacuumDens} we find
\begin{equation}
\dot{\rho}_{m} + 3H\xi\rho_m = 0    
\end{equation}
where, for convenience, we have defined
\begin{equation}
\xi \equiv \frac{1 -\nu}{1 - \frac{3}{4}\nu}.    
\end{equation}
We encode the deviations with respect to the standard model in terms of the effective parameter $\nueff\equiv\nu/4$:
\begin{equation}
\xi = 1-\nu_{\rm eff} + \mathcal{O}\left(\nu_{\rm eff}^2\right)\,.
\end{equation}
Having reached this point it is straightforward to find the expression for the matter energy densities:
\begin{equation}\label{eq:MassDensities}
\rho_m(a) = \rho^0_m{a^{-3\xi}}\,, \ \ \ \rho_{\rm cdm}(a) = \rho^{0}_m{a^{-3\xi}}  - \rho^0_b{a^{-3}} \,.
\end{equation}
As expected, by setting $\xi=1$ ($\nueff=0$) we recover the $\Lambda$CDM expressions. The small departure is precisely what gives allowance for a mild dynamical vacuum evolution:
\begin{align}  \label{Vacdensity}
\rv(a) &= \rvo + \left(\frac{1}{\xi} -1\right)\rho^0_m\left(a^{-3\xi} -1\right)\,.
\end{align}
The vacuum  becomes rigid if $\xi=1$ ($\nueff=0$).
\subsection{Type II RRVM}
Unlike the first kind of models presented, for type II models matter is conserved. However, vacuum can still evolve provided the gravitational coupling also evolves with the expansion: $G=G(H)$. Let us define an auxiliary variable  $\varphi=G_N/G $ -- in the manner of  a Brans-Dicke, without being really so. Notice that $\varphi\neq 1$ in the cosmological domain, but remains very close to it, see Tables 1 and 2. 
The modified Friedmann's equation for type-II model can be written as
\begin{equation}\label{eq:fried}
3H^2=\frac{8\pi G_N}{\varphi}\left[\rho_t+C_0+\frac{3\nu}{16\pi G_N}(2H^2+\dot{H})\right]\,,
\end{equation}
with $C_0=3c_0/(8\pi G_N)$.
The link between the dynamics of $\varphi$ and that of $\rv$ is given by the Bianchi identity: 
\begin{equation}\label{eq:Bianchi}
\frac{\dot{\varphi}}{\varphi}=\frac{\dot{\rho}_{\rm vac}}{\rho_t+\rv}\,,
\end{equation}
where $\rho_t $ is as before the total matter energy density and $\rv$ adopts exactly the same form as in \eqref{RRVM}. In the absence of an analytical expression for the vacuum energy density, we can show its approximate behaviour close to the present time by keeping only the terms linear in $\nu_{\rm eff}$ (recall that $|\nueff|\ll1$):
\begin{equation}\label{eq:VDEm}
\rv(a)=C_0(1+4\nueff)+\nueff\rho_m^{0}a^{-3}+\mathcal{O}(\nueff^2)\,.
\end{equation}
Again, for $\nueff=0$ the vacuum energy density is constant, but otherwise it shows a moderate dynamics of ${\cal O}(\nueff)$ as in the type I case \eqref{Vacdensity}. One can also show that  $\rv(a)\ll\rho_r(a)=\rho_r^0 a^{-4}$  for $a\ll 1$ and therefore the $\rho_{\rm vac}$ for the type II model does not perturb the normal thermal history (as in the type I model). Regarding the auxiliary variable $\varphi$, in the current epoch it exhibits a very mild evolution, almost logarithmic. 
\subsection{Threshold redshift scenario for type I models}
One possibility that has been explored in the literature is to consider that vacuum is a dynamical quantity only close to the present time and that remains constant for the rest of the cosmic history (see e.g. \cite{Salvatelli:2014zta,Martinelli:2019dau}). So, in this scenario, we keep deactivated the interaction between the vacuum energy density and the CDM until the late universe when dark energy becomes apparent. We denote the threshold value of the scale factor when the activation takes places by $\astar$. According to this scenario the vacuum energy density was constant prior to $a=\astar$ and it just started to evolve for $a>\astar$. It is important to remark that while $\rv$ is a continuous function, its derivative is not. We mimic such situation through a Heaviside step function  $\Theta(a-\astar)$. Therefore, we assume that in the range   $a<\astar$  (hence for $z> \zstar$) we have
\begin{align}
&\rho_{\rm cdm}(a) = \rho_{\rm cdm}(\astar)\left(\frac{a}{\astar}\right)^{-3},  \ \ \ \ \ \ \ \ \ \  \nonumber\\
&\rv (a) = \rv(\astar)=\text{const.} \ \ \ \ \ \ \ \ \ \ \ \ \ \   (a<\astar)\,, \label{Vacdensitystar}
\end{align}
where $\rho_{\rm cdm}(\astar)$ and  $\rv(\astar)$ are computed from \eqref{eq:MassDensities} and \eqref{Vacdensity}, respectively. In the complementary range, instead,  i.e. for  $a>\astar$   ($0<z<\zstar$) near our time, the original equations \eqref{eq:MassDensities} and \eqref{Vacdensity} are the ones considered. 
\newline
\newline
The threshold procedure is only applied within type I models with the main purpose of preserving the standard evolution law for the matter energy density when the redshift is sufficiently high. In fact, the threshold redshift value does not need to be very high and when it is fixed by optimization it turns out to be of order  $\zstar\simeq 1$.  An important consequence of  such threshold is that the cosmological physics during the  CMB epoch is exactly as in the $\CC$CDM. On the contrary for type II models there is still some evolution (very mild though) of $\rho_{\rm vac}$ at the CMB epoch, but the matter density follows the same law as in the standard model.
\section{Cosmological perturbations}
In order to perform a complete analysis of the different models under study we need to consider the evolution of the perturbed cosmological quantities throughout the cosmic history. We refer the reader to the references \cite{SolaPeracaula:2020vpg, SolaPeracaula:2016qlq, SolaPeracaula:2017esw, Gomez-Valent:2018nib} to know for the details of the perturbation equations since here we just display some basic equations. We consider the perturbed, spatially flat, FLRW metric $ds^2=a^2(\eta)[-d\eta^2+(\delta_{ij}+h_{ij})dx^idx^j]$, in which $h_{ij}$ represents the metric fluctuations which are coupled to the matter density perturbations $\delta_m = \delta\rho_m/\rho_m$ and $d\eta = dt/a$ is the conformal time. 
In the case of the BD model, at deep subhorizon scales, the differential equation for the matter density contrast is: 
\begin{equation}
\delta^{\prime\prime}_m + \mathcal{H}\delta^{\prime}_m - 4\pi{G_{\rm eff}(\bar{\varphi})}{a^2}\bar{\rho}_m\delta_m = 0,     
\end{equation}
where here $()^\prime \equiv d()/d \eta$ and $\mathcal{H} = aH$. In this section when we add a bar over a quantity we indicate that it is a background quantity. We have employed the definition:
\begin{equation}
G_{\rm eff} = \frac{G_N}{\bar{\varphi}}\left(\frac{4 + 2\omega_{\rm BD}}{3 + 2\omega_{\rm BD}}\right),   
\end{equation}
which represents the effective coupling that modifies the Poisson term of the perturbed equation with respect to the standard model. 
\newline
\newline 
In the case of the type I RRVM since baryons do not interact with the time evolving vacuum energy density their perturbed conservation equations are not directly affected. On the other hand the equation for the CDM takes the following form: 
\begin{equation}
\delta^{\prime}_{\rm cdm} + \frac{h^{\prime}}{2} - \frac{\rho^{\prime}_{\rm vac}}{\rho_{\rm  cdm}}\delta_{\rm cdm} = 0,    
\end{equation}
which is obviously affected by the dynamics of the vacuum in a nontrivial way. We will present the details of the perturbation equations for the type II RRVM elsewhere. 
\section{Data and Methodology}
We fit the BD-$\Lambda$CDM, the different RVM's together with the standard model to the wealth of the cosmological data compiled from distant type Ia supernovae (SNIa), baryonic acoustic oscillations (BAO), different values of the Hubble function $H(z_i)$, the large scale structure (LSS) formation data embodied in the $f(z_i)\sigma_8(z_i)$ observable and the CMB data from the Planck satellite, see the references in the caption of Table 1. 
In order to compare the theoretical predictions with the available cosmological data we define the total $\chi^2$ function as:
\begin{equation}
\chi^2_{\rm tot} = \chi^2_{\rm SNIa} + \chi^2_{\rm BAO} + \chi^2_H + \chi^2_{f\sigma_8} + \chi^2_{\rm CMB}.   
\end{equation}
The above terms are defined in the standard way from the data including the corresponding covariance matrices. In particular, the $\chi^2_H$ part may contain or not the $H_0$ value measured by \cite{Reid:2019tiq} depending on the setup indicated in the tables. 
To obtain the posterior distributions and the corresponding constraints for the dataset described above we have run the Monte Carlo sampler \texttt{MontePython} \cite{Audren:2012wb} together with the Einstein-Boltzmann system solver \texttt{CLASS} \cite{Blas:2011rf}. The latter has been properly modified in order to implement the background and the linear perturbation equations for the different models. 
\section{Discussions and conclusions}
The main fitting results of our analysis are displayed in Tables 1 and 2 and in Figure 1. In the tables we compare, for two different datasets denoted as ``Baseline" and ``Baseline+$H_0$'' (see the corresponding captions for more details), the concordance $\Lambda$CDM model with different RRVM's and with the BD-$\Lambda$CDM model. 
\newline
In order to make a fair comparison between models with a different numbers of free parameters we employ the Deviance Information Criterion (DIC) \cite{DIC}: $\Delta{\rm DIC} = {\rm DIC}_{\rm \Lambda{CDM}}- {\rm DIC}_{\rm X}$, representing X one of the nonstandard models under study. The DIC is defined as
\begin{equation}
{\rm DIC}=\chi^2(\overline{\theta})+2p_D\,.
\end{equation}
Here $p_D=\overline{\chi^2}-\chi^2(\overline{\theta})$ is the effective number of parameters of the model, and $\overline{\chi^2}$ and $\overline{\theta}$ the mean of the overall $\chi^2$ distribution and the parameters, respectively. As it can be seen from the definition of $\Delta{\rm DIC}$, if $\Delta{\rm DIC} < 0$, means that the $\Lambda$CDM fits better the cosmological data, whereas for $\Delta{\rm DIC} > 0$ is the other way around. \newline
\newline
%
%
\begin{figure*}
\centering
\includegraphics[angle=0,width=0.9\linewidth]{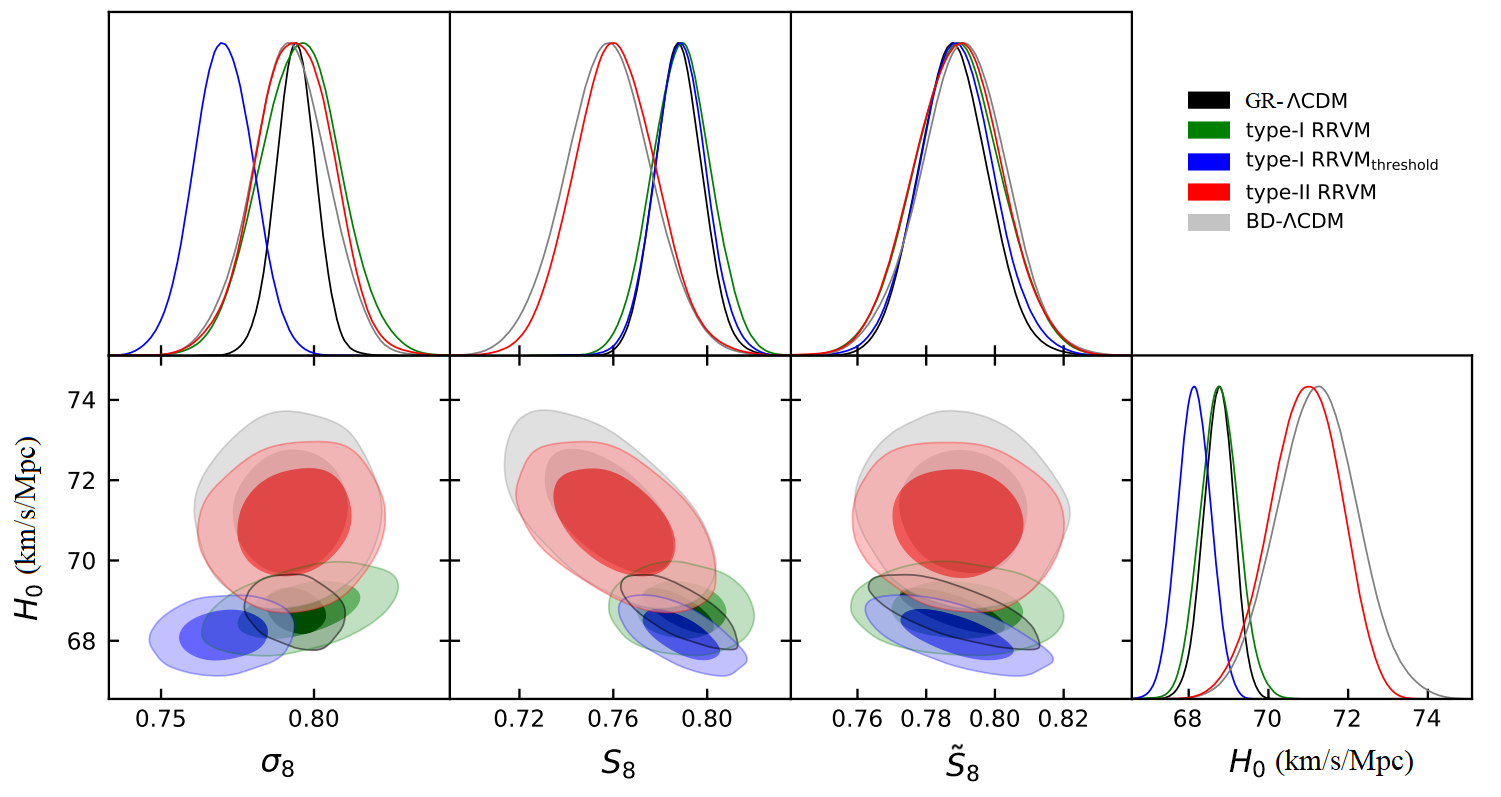}
\caption{\label{fig:XCDMEvolution}%
The contour plots at $1\sigma$ and $2\sigma$ confidence level in the $H_0$-$\sigma_8,S_8,\tilde{S}_8$ planes and the corresponding one-dimensional posteriors for the {GR- and BD- $\Lambda$CDM and the RRVM's} obtained from the fitting analyses with our Baseline+$H_0$ data set. The type II model manifestly alleviates the $H_0$ tension without spoiling the $\sigma_8$ one, whereas the type I model with threshold redshift $\zstar\simeq 1$ can fully solve the latter but cannot address the former.
}
\end{figure*}
%
%
Taking a look at the first table, it can be appreciated that a mild dynamics of the vacuum is very much welcome, specially in the case where the time evolution of $\rho_{\rm vac}$ is activated close to the moment when the contribution of this quantity becomes relevant, namely at $z\simeq 1$. It is in this particular scenario where the impact on the description of the cosmological data string SNIa+BAO+$H(z)$+LSS+CMB becomes extraordinary significant on statistical terms, since $\Delta{\rm AIC}> +10$ which means that a very strong evidence, in favor of this model, is found, when it is compared with the $\Lambda$CDM model. At very high redshift the physics of the type I RRVM with a threshold remain basically unaltered with respect to the $\Lambda$CDM, however, the activation of the dynamics of the vacuum at $z_{*}\simeq 1$ allows to suppress an exceedingly amount of structure in the universe, thus leading to a better description of the $f(z)\sigma_8(z)$ data set.  On the contrary, non of the other models beyond the standard one is preferred by the cosmological data, and the DIC indicates moderate evidence against them ($-5<{\rm DIC} < -2$). We would like to remark that even if the RRVM's under study are not favoured by the data, in all cases the values of $S_8\equiv \sigma_8(0)\sqrt{\Omega^0_m/0.3}$ (or the analogous observable $\tilde{S}_8 \equiv S_8/\sqrt{\varphi_0}$ for the type II RRVM and the BD-$\Lambda$CDM models) remain compatible with recent weak lensing and galaxy clustering measurements \cite{Heymans:2020gsg}, hence smoothing out the $\sigma_8$-tension.  Just by including the $H_0$ prior in the data set the results change in a very significant way as it can be observed from Table 2. The $\Lambda$CDM model, as well as, the two versions of the type I RRVM model are not able to accommodate high values of the $H_0$ parameter, hence, all of them show their inability to alleviate the $H_0$-tension. However, the DIC still decides very strongly in the case of the type I RRVM with threshold. On the other hand, the type II RRVM and the BD-$\Lambda$CDM model seem to have no problem to fit the $H_0$ prior since in both cases the fitting value for this parameter is $\simeq 71$km/s/Mpc. To fully appreciate how the cosmological models deal with the $H_0$ and the $\sigma_8$ tensions simultaneously, we have included Figure 1, where we have depicted the contour plots for the different models in the  $H_0$-$\sigma_8,S_8,\tilde{S}_8$ planes of the parameter space. As mentioned before, the RRVM's are able to keep the values of the different parameters related with the structure formation ($\sigma_8,S_8,\tilde{S}_8$) at an intermediate value between Planck measurements \cite{Planck:2018vyg} and cosmic shear data \cite{Heymans:2020gsg}. Remarkably enough, in the case of the type II RRVM and the BD-$\Lambda$CDM models, the contour lines in Figure 1 show a preference for relatively high values of $H_0$ (the tension is lowered at $\sim 1.6\sigma$) while keeping $\sigma_8\equiv \sigma_8(0)$ in the low value range (the tension remains at $\sim 0.4\sigma$), thus smoothing out both tensions at a time. We conclude that the models studied in this paper provide very interesting alternatives, either considering a time-evolving vacuum energy density or by extending the paradigm of General Relativity, to alleviate the $H_0$ and the $\sigma_8$ tensions. The better performance of the aforementioned models is reconfirmed by the Deviance Information Criterion, which points out that in some cases a very strong evidence in favor of a nonstandard model is found. 
\section*{Acknowledgments}
I would like to thank J. Sol\`a Peracaula and A. G\'omez-Valent for inviting me to speak in the CM3 parallel sesion ``Status of the $H_0$ and $\sigma_8$ Tensions: Theoretical Models and Model-Independent Constraints of the MG16 Marcel Grossman Virtual Conference, July 5-10 2021. We also would like to thank the organizers of the MG16 for their great effort organizing such a great event. JdCP is supported by a  FPI fellowship associated to the project FPA2016-76005-C2-1-P. JSP and CMP are partially supported by projects  FPA2016-76005-C2-1-P (MINECO), 2017-SGR-929 (Generalitat de Catalunya) and MDM-2014-0369 (ICCUB). CMP is partially supported by the fellowship 2019 FI-B 00351. AGV is funded by the INFN -- Project number 22425/2020.

\bibliographystyle{ws-procs961x669}
\bibliography{ws-pro-sample}

\begin{thebibliography}{10}

\bibitem{Riess:1998cb}
A.~G. Riess {\em et~al.}, Observational evidence from supernovae for an
  accelerating universe and a cosmological constant, {\em Astron. J.} {\bf
  116}, 1009  (1998), arXiv:astro-ph/9805201.

\bibitem{Perlmutter:1998np}
S.~Perlmutter {\em et~al.}, {Measurements of $\Omega$ and $\Lambda$ from 42
  high redshift supernovae}, {\em Astrophys. J.} {\bf 517}, 565  (1999),
  arXiv:astro-ph/9812133.

\bibitem{Einstein:1915ca}
A.~Einstein, {The Field Equations of Gravitation}, {\em Sitzungsber. Preuss.
  Akad. Wiss. Berlin (Math. Phys. )} {\bf 1915}, 844  (1915).

\bibitem{Zeldovich:1967}
Y.~Zel'dovich, {Cosmological constant and elementary particles}, {\em Sov.
  Phys. JETP. Lett.} {\bf 6}, p. 3167  (1967).

\bibitem{Zeldovich:1968}
Y.~Zel'dovich, {The cosmological constant and the theory of elementary
  particles}, {\em Sov. Phys. Ups.} {\bf 11}, p. 381  (1968).

\bibitem{Weinberg:1988cp}
S.~Weinberg, {The Cosmological Constant Problem}, {\em Rev. Mod. Phys.} {\bf
  61}, 1  (1989).

\bibitem{Peebles:2002gy}
P.~J.~E. Peebles and B.~Ratra, {The Cosmological Constant and Dark Energy},
  {\em Rev. Mod. Phys.} {\bf 75}, 559  (2003), arXiv:astro-ph/0207347.

\bibitem{Padmanabhan:2002ji}
T.~Padmanabhan, {Cosmological constant: The Weight of the vacuum}, {\em Phys.
  Rept.} {\bf 380}, 235  (2003), arXiv:hep-th/0212290.

\bibitem{Copeland:2006wr}
E.~J. Copeland, M.~Sami and S.~Tsujikawa, {Dynamics of dark energy}, {\em Int.\
  J.\ Mod.\ Phys.\ D} {\bf 15}, 1753  (2006), arXiv:hep-th/0603057.

\bibitem{Planck:2018vyg}
N.~Aghanim {\em et~al.}, {Planck 2018 results. VI. Cosmological parameters},
  {\em Astron. Astrophys.} {\bf 641}, p.~A6  (2020), [Erratum:
  Astron.Astrophys. 652, C4 (2021)].

\bibitem{Reid:2019tiq}
M.~J. Reid, D.~W. Pesce and A.~G. Riess, {An Improved Distance to NGC 4258 and
  its Implications for the Hubble Constant}, {\em Astrophys. J.} {\bf 886}, p.
  L27  (2019), arXiv:1908.05625.

\bibitem{Macaulay:2013swa}
E.~Macaulay, I.~K. Wehus and H.~K. Eriksen, {Lower Growth Rate from Recent
  Redshift Space Distortion Measurements than Expected from Planck}, {\em Phys.
  Rev. Lett.} {\bf 111}, p. 161301  (2013), arXiv:1303.6583.

\bibitem{Gomez-Valent:2021cbe}
A.~G\'omez-Valent, Z.~Zheng, L.~Amendola, V.~Pettorino and C.~Wetterich, {Early
  dark energy in the pre- and post-recombination epochs} (7 2021).

\bibitem{Benevento:2020fev}
G.~Benevento, W.~Hu and M.~Raveri, {Can Late Dark Energy Transitions Raise the
  Hubble constant?}, {\em Phys. Rev. D} {\bf 101}, p. 103517  (2020),
  arXiv:2002.11707.

\bibitem{Peebles:1987ek}
P.~J.~E. Peebles and B.~Ratra, {Cosmology with a Time Variable Cosmological
  Constant}, {\em Astrophys. J.} {\bf 325}, p. L17  (1988).

\bibitem{Ratra:1987rm}
B.~Ratra and P.~J.~E. Peebles, {Cosmological Consequences of a Rolling
  Homogeneous Scalar Field}, {\em Phys. Rev.} {\bf D37}, p. 3406  (1988).

\bibitem{Peccei:1987mm}
R.~D. Peccei, J.~Sol{\`{a}} and C.~Wetterich, {Adjusting the Cosmological
  Constant Dynamically: Cosmons and a New Force Weaker Than Gravity}, {\em
  Phys. Lett.} {\bf B195}, 183  (1987).

\bibitem{Wetterich:1987fm}
C.~Wetterich, {Cosmology and the Fate of Dilatation Symmetry}, {\em Nucl.
  Phys.} {\bf B302}, 668  (1988), arXiv:1711.03844.

\bibitem{Wetterich:1994bg}
C.~Wetterich, {The Cosmon model for an asymptotically vanishing time dependent
  cosmological 'constant'}, {\em Astron. Astrophys.} {\bf 301}, 321  (1995),
  arXiv:hep-th/9408025.

\bibitem{Amendola:1999er}
L.~Amendola, {Coupled quintessence}, {\em Phys. Rev.} {\bf D62}, p. 043511
  (2000), arXiv:astro-ph/9908023.

\bibitem{Gomez-Valent:2020mqn}
A.~Gómez-Valent, V.~Pettorino and L.~Amendola, {Update on coupled dark energy
  and the $H_0$ tension}, {\em Phys. Rev. D} {\bf 101}, p. 123513  (2020),
  arXiv:2004.00610.

\bibitem{Sola:2016jky}
J.~Sol\`a, A.~G\'omez-Valent and J.~de~Cruz~P\'erez, {First evidence of running
  cosmic vacuum: challenging the concordance model}, {\em Astrophys. J.} {\bf
  836}, p.~43  (2017).

\bibitem{SolaPeracaula:2017esw}
J.~Sol\`a~Peracaula, J.~de~Cruz~P\'erez and A.~G\'omez-Valent, {Possible
  signals of vacuum dynamics in the Universe}, {\em Mon. Not. Roy. Astron.
  Soc.} {\bf 478}, 4357  (2018).

\bibitem{Sola:2017znb}
J.~Sol{\`{a}}, A.~G{\'{o}}mez-Valent and J.~de~Cruz~P{\'{e}}rez, {The $H_0$
  tension in light of vacuum dynamics in the Universe}, {\em Phys. Lett.} {\bf
  B774}, 317  (2017), arXiv:1705.06723.

\bibitem{Gomez-Valent:2018nib}
A.~G{\'{o}}mez-Valent and J.~Sol{\`{a}}~Peracaula, {Density perturbations for
  running vacuum: a successful approach to structure formation and to the
  $\sigma_8$-tension}, {\em Mon. Not. Roy. Astron. Soc.} {\bf 478}, 126
  (2018), arXiv:1801.08501.

\bibitem{Sola:2018sjf}
J.~Sol{\`{a}}~Peracaula, A.~G{\'{o}}mez-Valent and J.~de~Cruz~P{\'{e}}rez,
  {Signs of Dynamical Dark Energy in Current Observations}, {\em Phys. Dark
  Univ.} {\bf 25}, p. 100311  (2019), arXiv:1811.03505.

\bibitem{SolaPeracaula:2021gxi}
J.~Sol\`a~Peracaula, A.~G\'omez-Valent, J.~de~Cruz~P\'erez and
  C.~Moreno-Pulido, {Running vacuum against the $H_0$ and $\sigma_8$ tensions},
  {\em EPL} {\bf 134}, p. 19001  (2021).

\bibitem{BransDicke1961}
C.~Brans and R.~Dicke, Mach's principle and a relativistic theory of
  gravitation, {\em Phys. Rev} {\bf 124}, p. 925  (1961).

\bibitem{SolaPeracaula:2019zsl}
J.~Sol\`a~Peracaula, A.~G\'omez-Valent, J.~de~Cruz~P\'erez and
  C.~Moreno-Pulido, {Brans\textendash{}Dicke Gravity with a Cosmological
  Constant Smoothes Out $\Lambda$CDM Tensions}, {\em Astrophys. J. Lett.} {\bf
  886}, p.~L6  (2019).

\bibitem{SolaPeracaula:2020vpg}
J.~Sol\`a~Peracaula, A.~G\'omez-Valent, J.~de~Cruz~P\'erez and
  C.~Moreno-Pulido, {Brans\textendash{}Dicke cosmology with a $\Lambda$-term: a
  possible solution to $\Lambda$CDM tensions}, {\em Class. Quant. Grav.} {\bf
  37}, p. 245003  (2020).

\bibitem{MTW:1974}
C.~W. Misner, K.~S. Thorn and J.~A. Wheeler, {\em Gravitation} (Freeman, San
  Francisco, 1974).

\bibitem{Ooba:2017ukj}
J.~Ooba, B.~Ratra and N.~Sugiyama, {Planck 2015 Constraints on the Non-flat
  $\Lambda$CDM Inflation Model}, {\em Astrophys. J.} {\bf 864}, p.~80  (2018).

\bibitem{Ooba:2017lng}
J.~Ooba, B.~Ratra and N.~Sugiyama, {Planck 2015 Constraints on the Nonflat
  $\phi$CDM Inflation Model}, {\em Astrophys. J.} {\bf 866}, p.~68  (2018).

\bibitem{Park:2018bwy}
C.-G. Park and B.~Ratra, {Observational constraints on the tilted flat-XCDM and
  the untilted nonflat XCDM dynamical dark energy inflation parameterizations},
  {\em Astrophys. Space Sci.} {\bf 364}, p.~82  (2019).

\bibitem{Park:2018fxx}
C.-G. Park and B.~Ratra, {Observational constraints on the tilted
  spatially-flat and the untilted nonflat $\phi$CDM dynamical dark energy
  inflation models}, {\em Astrophys. J.} {\bf 868}, p.~83  (2018).

\bibitem{Park:2018tgj}
C.-G. Park and B.~Ratra, {Measuring the Hubble constant and spatial curvature
  from supernova apparent magnitude, baryon acoustic oscillation, and Hubble
  parameter data}, {\em Astrophys. Space Sci.} {\bf 364}, p. 134  (2019).

\bibitem{Park:2019emi}
C.-G. Park and B.~Ratra, {Using SPT polarization, $Planck$ 2015, and non-CMB
  data to constrain tilted spatially-flat and untilted nonflat $\Lambda$CDM ,
  XCDM, and $\phi$CDM dark energy inflation cosmologies}, {\em Phys. Rev. D}
  {\bf 101}, p. 083508  (2020).

\bibitem{Ryan:2019uor}
J.~Ryan, Y.~Chen and B.~Ratra, {Baryon acoustic oscillation, Hubble parameter,
  and angular size measurement constraints on the Hubble constant, dark energy
  dynamics, and spatial curvature}, {\em Mon. Not. Roy. Astron. Soc.} {\bf
  488}, 3844  (2019).

\bibitem{Khadka:2020vlh}
N.~Khadka and B.~Ratra, {Using quasar X-ray and UV flux measurements to
  constrain cosmological model parameters}, {\em Mon. Not. Roy. Astron. Soc.}
  {\bf 497}, 263  (2020), arXiv:2004.09979.

\bibitem{Khadka:2020whe}
N.~Khadka and B.~Ratra, {Quasar X-ray and UV flux, baryon acoustic oscillation,
  and Hubble parameter measurement constraints on cosmological model
  parameters}, {\em Mon. Not. Roy. Astron. Soc.} {\bf 492}, 4456  (2020).

\bibitem{Cao:2020evz}
S.~Cao, J.~Ryan, N.~Khadka and B.~Ratra, {Cosmological constraints from higher
  redshift gamma-ray burst, H ii starburst galaxy, and quasar (and other)
  data}, {\em Mon. Not. Roy. Astron. Soc.} {\bf 501}, 1520  (2021).

\bibitem{Cao:2021ldv}
S.~Cao, J.~Ryan and B.~Ratra, {Using Pantheon and DES supernova, baryon
  acoustic oscillation, and Hubble parameter data to constrain the Hubble
  constant, dark energy dynamics, and spatial curvature}, {\em Mon. Not. Roy.
  Astron. Soc.} {\bf 504}, 300  (2021).

\bibitem{Khadka:2021xcc}
N.~Khadka and B.~Ratra, {Do quasar X-ray and UV flux measurements provide a
  useful test of cosmological models?} (7 2021).

\bibitem{Cao:2021cix}
S.~Cao, J.~Ryan and B.~Ratra, {Cosmological constraints from HII starburst
  galaxy, quasar angular size, and other measurements} (9 2021).

\bibitem{Faraoni:1998yq}
V.~Faraoni, {The $\omega\to\infty$ limit of Brans Dicke theory}, {\em Phys.
  Lett.} {\bf A245}, 26  (1998), arXiv:gr-qc/9805057.

\bibitem{Faraoni:1999yp}
V.~Faraoni, {Illusions of general relativity in Brans-Dicke gravity}, {\em
  Phys. Rev.} {\bf D59}, p. 084021  (1999), arXiv:gr-qc/9902083.

\bibitem{Scolnic:2017caz}
D.~M. Scolnic {\em et~al.}, {The Complete Light-curve Sample of
  Spectroscopically Confirmed SNe Ia from Pan-STARRS1 and Cosmological
  Constraints from the Combined Pantheon Sample}, {\em Astrophys. J.} {\bf
  859}, p. 101  (2018), arXiv:1710.00845.

\bibitem{Carter:2018vce}
P.~Carter, F.~Beutler, W.~J. Percival, C.~Blake, J.~Koda and A.~J. Ross, {Low
  Redshift Baryon Acoustic Oscillation Measurement from the Reconstructed
  6-degree Field Galaxy Survey}, {\em Mon. Not. Roy. Astron. Soc.} {\bf 481},
  2371  (2018), arXiv:1803.01746.

\bibitem{Kazin:2014qga}
E.~A. Kazin {\em et~al.}, {The WiggleZ Dark Energy Survey: improved distance
  measurements to z = 1 with reconstruction of the baryonic acoustic feature},
  {\em Mon. Not. Roy. Astron. Soc.} {\bf 441}, 3524  (2014), arXiv:1401.0358.

\bibitem{Gil-Marin:2016wya}
H.~Gil-Mar{\'{i}}n, W.~J. Percival, L.~Verde, J.~R. Brownstein, C.-H. Chuang,
  F.-S. Kitaura, S.~A. Rodr{\'{i}}guez-Torres and M.~D. Olmstead, {The
  clustering of galaxies in the SDSS-III Baryon Oscillation Spectroscopic
  Survey: RSD measurement from the power spectrum and bispectrum of the DR12
  BOSS galaxies}, {\em Mon. Not. Roy. Astron. Soc.} {\bf 465}, 1757  (2017),
  arXiv:1606.00439.

\bibitem{Abbott:2017wcz}
T.~M.~C. Abbott {\em et~al.}, {Dark Energy Survey Year 1 Results: Measurement
  of the Baryon Acoustic Oscillation scale in the distribution of galaxies to
  redshift 1}, {\em Mon. Not. Roy. Astron. Soc.} {\bf 483}, 4866  (2019),
  arXiv:1712.06209.

\bibitem{Neveux:2020voa}
R.~Neveux {\em et~al.}, {The completed SDSS-IV extended Baryon Oscillation
  Spectroscopic Survey: BAO and RSD measurements from the anisotropic power
  spectrum of the quasar sample between redshift 0.8 and 2.2}, {\em Mon. Not.
  Roy. Astron. Soc.} {\bf 499}, 210  (2020).

\bibitem{duMasdesBourboux:2020pck}
H.~du~Mas~des Bourboux {\em et~al.}, {The Completed SDSS-IV Extended Baryon
  Oscillation Spectroscopic Survey: Baryon Acoustic Oscillations with
  Ly\ensuremath{\alpha} Forests}, {\em Astrophys. J.} {\bf 901}, p. 153
  (2020).

\bibitem{Jimenez:2003iv}
R.~Jim{\'{e}}nez, L.~Verde, T.~Treu and D.~Stern, {Constraints on the equation
  of state of dark energy and the Hubble constant from stellar ages and the
  CMB}, {\em Astrophys. J.} {\bf 593}, 622  (2003), arXiv:astro-ph/0302560.

\bibitem{Simon:2004tf}
J.~Simon, L.~Verde and R.~Jimenez, {Constraints on the redshift dependence of
  the dark energy potential}, {\em Phys. Rev.} {\bf D71}, p. 123001  (2005),
  arXiv:astro-ph/0412269.

\bibitem{Stern:2009ep}
D.~Stern, R.~Jim{\'{e}}nez, L.~Verde, M.~Kamionkowski and S.~A. Stanford,
  {Cosmic Chronometers: Constraining the Equation of State of Dark Energy. I:
  H(z) Measurements}, {\em JCAP} {\bf 1002}, p. 008  (2010), arXiv:0907.3149.

\bibitem{Moresco:2012jh}
M.~Moresco {\em et~al.}, {Improved constraints on the expansion rate of the
  Universe up to z~1.1 from the spectroscopic evolution of cosmic
  chronometers}, {\em JCAP} {\bf 1208}, p. 006  (2012), arXiv:1201.3609.

\bibitem{Zhang:2012mp}
C.~Zhang, H.~Zhang, S.~Yuan, T.-J. Zhang and Y.-C. Sun, {Four new observational
  $H(z)$ data from luminous red galaxies in the Sloan Digital Sky Survey data
  release seven}, {\em Res. Astron. Astrophys.} {\bf 14}, 1221  (2014),
  arXiv:1207.4541.

\bibitem{Moresco:2015cya}
M.~Moresco, {Raising the bar: new constraints on the Hubble parameter with
  cosmic chronometers at $z \sim 2$}, {\em Mon. Not. Roy. Astron. Soc.} {\bf
  450}, L16  (2015), arXiv:1503.01116.

\bibitem{Moresco:2016mzx}
M.~Moresco, L.~Pozzetti, A.~Cimatti, R.~Jim{\'{e}}nez, C.~Maraston, L.~Verde,
  D.~Thomas, A.~Citro, R.~Tojeiro and D.~Wilkinson, {A 6\% measurement of the
  Hubble parameter at $z\sim0.45$: direct evidence of the epoch of cosmic
  re-acceleration}, {\em JCAP} {\bf 1605}, p. 014  (2016), arXiv:1601.01701.

\bibitem{Ratsimbazafy:2017vga}
A.~L. Ratsimbazafy, S.~I. Loubser, S.~M. Crawford, C.~M. Cress, B.~A. Bassett,
  R.~C. Nichol and P.~V{\"{a}}is{\"{a}}nen, {Age-dating Luminous Red Galaxies
  observed with the Southern African Large Telescope}, {\em Mon. Not. Roy.
  Astron. Soc.} {\bf 467}, 3239  (2017), arXiv:1702.00418.

\bibitem{Said:2020epb}
K.~Said, M.~Colless, C.~Magoulas, J.~R. Lucey and M.~J. Hudson, {Joint analysis
  of 6dFGS and SDSS peculiar velocities for the growth rate of cosmic structure
  and tests of gravity}, {\em Mon. Not. Roy. Astron. Soc.} {\bf 497}, 1275
  (2020).

\bibitem{Simpson:2015yfa}
F.~Simpson, C.~Blake, J.~A. Peacock, I.~Baldry, J.~Bland-Hawthorn, A.~Heavens,
  C.~Heymans, J.~Loveday and P.~Norberg, {Galaxy and mass assembly: Redshift
  space distortions from the clipped galaxy field}, {\em Phys. Rev.} {\bf D93},
  p. 023525  (2016), arXiv:1505.03865.

\bibitem{Blake:2013nif}
C.~Blake {\em et~al.}, {Galaxy And Mass Assembly (GAMA): improved cosmic growth
  measurements using multiple tracers of large-scale structure}, {\em Mon. Not.
  Roy. Astron. Soc.} {\bf 436}, p. 3089  (2013), arXiv:1309.5556.

\bibitem{Mohammad:2018mdy}
F.~G. Mohammad {\em et~al.}, {The VIMOS Public Extragalactic Redshift Survey
  (VIPERS): Unbiased clustering estimate with VIPERS slit assignment}, {\em
  Astron. Astrophys.} {\bf 619}, p. A17  (2018), arXiv:1807.05999.

\bibitem{Okumura:2015lvp}
T.~Okumura {\em et~al.}, {The Subaru FMOS galaxy redshift survey (FastSound).
  IV. New constraint on gravity theory from redshift space distortions at
  $z\sim 1.4$}, {\em Publ. Astron. Soc. Jap.} {\bf 68}, p.~38  (2016),
  arXiv:1511.08083.

\bibitem{Peracaula:2018dkg}
J.~Solà~Peracaula, {Brans–Dicke gravity: From Higgs physics to (dynamical)
  dark energy}, {\em Int. J. Mod. Phys.} {\bf D27}, p. 1847029  (2018),
  arXiv:1805.09810.

\bibitem{Perez:2018qgw}
J.~de~Cruz~P{\'{e}}rez and J.~Sol{\`{a}}~Peracaula, {Brans–Dicke cosmology
  mimicking running vacuum}, {\em Mod. Phys. Lett.} {\bf A33}, p. 1850228
  (2018), arXiv:1809.03329.

\bibitem{Banerjee:2000mj}
N.~Banerjee and D.~Pavon, {Cosmic acceleration without quintessence}, {\em
  Phys. Rev. D} {\bf 63}, p. 043504  (2001).

\bibitem{Banerjee:2000gt}
N.~Banerjee and D.~Pavon, {A Quintessence scalar field in Brans-Dicke theory},
  {\em Class. Quant. Grav.} {\bf 18}, p. 593  (2001), arXiv:gr-qc/0012098.

\bibitem{Sola:2013gha}
J.~Sol{\`{a}}, {Cosmological constant and vacuum energy: old and new ideas},
  {\em J. Phys. Conf. Ser.} {\bf 453}, p. 012015  (2013), arXiv:1306.1527.

\bibitem{Sola:2015rra}
J.~Sol\`a and A.~G\'omez-Valent, {The $\bar{\Lambda}{\rm CDM}$ cosmology: From
  inflation to dark energy through running \ensuremath{\Lambda}}, {\em Int. J.
  Mod. Phys. D} {\bf 24}, p. 1541003  (2015).

\bibitem{Gomez-Valent:2017tkh}
A.~Gómez-Valent, {Vacuum energy in Quantum Field Theory and Cosmology}, PhD
  thesis, ICC, Barcelona U.2017.
\newblock arXiv:1710.01978.

\bibitem{Moreno-Pulido:2020anb}
C.~Moreno-Pulido and J.~Sol\`a, {Running vacuum in quantum field theory in
  curved spacetime: renormalizing $\rho_{vac}$ without $\sim m^4$ terms}, {\em
  Eur. Phys. J. C} {\bf 80}, p. 692  (2020), arXiv:2005.03164.

\bibitem{Sola:2007sv}
J.~Solà, {Dark energy: A Quantum fossil from the inflationary Universe?}, {\em
  J. Phys.} {\bf A41}, p. 164066  (2008), arXiv:0710.4151.

\bibitem{Lima:2013dmf}
J.~A.~S. Lima, S.~Basilakos and J.~Sola, {Expansion History with Decaying
  Vacuum: A Complete Cosmological Scenario}, {\em Mon. Not. Roy. Astron. Soc.}
  {\bf 431}, 923  (2013).

\bibitem{Sola:2015csa}
J.~Sol\`a, {The cosmological constant and entropy problems: mysteries of the
  present with profound roots in the past}, {\em Int. J. Mod. Phys. D} {\bf
  24}, p. 1544027  (2015).

\bibitem{Blas:2011rf}
D.~Blas, J.~Lesgourgues and T.~Tram, {The Cosmic Linear Anisotropy Solving
  System (CLASS) II: Approximation schemes}, {\em JCAP} {\bf 1107}, p. 034
  (2011), arXiv:1104.2933.

\bibitem{Salvatelli:2014zta}
V.~Salvatelli, N.~Said, M.~Bruni, A.~Melchiorri and D.~Wands, {Indications of a
  late-time interaction in the dark sector}, {\em Phys. Rev. Lett.} {\bf 113},
  p. 181301  (2014), arXiv:1406.7297.

\bibitem{Martinelli:2019dau}
M.~Martinelli, N.~B. Hogg, S.~Peirone, M.~Bruni and D.~Wands, {Constraints on
  the interacting vacuum–geodesic CDM scenario}, {\em Mon. Not. Roy. Astron.
  Soc.} {\bf 488}, 3423  (2019), arXiv:1902.10694.

\bibitem{SolaPeracaula:2016qlq}
J.~Sol\`a~Peracaula, J.~de~Cruz~P\'erez and A.~G\'omez-Valent, {Dynamical dark
  energy vs. $\Lambda$ = const in light of observations}, {\em EPL} {\bf 121},
  p. 39001  (2018).

\bibitem{Audren:2012wb}
B.~Audren, J.~Lesgourgues, K.~Benabed and S.~Prunet, {Conservative Constraints
  on Early Cosmology: an illustration of the Monte Python cosmological
  parameter inference code}, {\em JCAP} {\bf 1302}, p. 001  (2013),
  arXiv:1210.7183.

\bibitem{DIC}
D.~J. Spiegelhalter, N.~G. Best, B.~P. Carlin and A.~van~der Linde, Bayesian
  measures of model complexity and fit, {\em J. Roy. Stat. Soc.} {\bf 64}, p.
  583  (2002).

\bibitem{Heymans:2020gsg}
C.~Heymans {\em et~al.}, {KiDS-1000 Cosmology: Multi-probe weak gravitational
  lensing and spectroscopic galaxy clustering constraints}, {\em Astron.
  Astrophys.} {\bf 646}, p. A140  (2021).

\end{thebibliography}

\end{document}